# Novel gas-doping technique for local spectroscopic measurements in pulsed-power systems

R. Arad,[a)] L. Ding,[b)] and Y. Maron
*Faculty of Physics, Weizmann Institute of Science, 76100 Rehovot, Israel*

A novel method for doping plasmas in pulsed-power experiments with gaseous elements has been developed. A fast gas valve, a nozzle, and a skimmer are used to generate an ultrasonic gas beam that is injected into a planar-geometry microsecond plasma-opening switch (POS). An array of ionization probes with relatively high spatial and temporal resolutions was developed for diagnosing the absolute injected-gas density and its spatial profile. The properties of the gas column were also studied using spectroscopy of line emission that results from the interaction of the doped gas with the POS prefilled plasma. The doped column is found to have a width of $\approx 1$ cm and a density of $(0.8-1.7) \times 10^{14}$ cm$^{-3}$. Observations of characteristic emission lines from the doped atoms and their ions allow for various spectroscopic measurements, such as the magnetic field from Zeeman splitting and the ion velocity distributions from Doppler shifts, that are local in three dimensions. It is shown that this gas doping technique can also be used to study proton-dominated plasmas that cannot be studied with simple emission spectroscopy due to the lack of light emitting ions. The variety of gases used with this method, together with the small valve dimensions and its fast opening, make it potentially useful for broad diagnostics of various short-duration plasma experiments. © *1998 American Institute of Physics.* [S0034-6748(98)03503-5]

## I. INTRODUCTION

The investigations of pulsed-power systems such as electron-beam diodes,[1] ion diodes,[2] z pinches,[3] and plasma opening switches[4–6] (POS) requires the use of nonperturbing techniques with high spatial and temporal resolutions. Methods based on interferometry[7] and emission spectroscopy,[8] highly advantageous for this purpose, suffer from the ambiguity caused by the measurement integration along the line of sight. For spectroscopic measurements, spatial resolution along the line of sight can be obtained by locally doping the plasma with selected species. To this end, we developed various doping techniques such as doping dielectric electrodes undergoing flashover[9] and laser evaporation of coated electrodes.[10] Here, we report on the development of molecular beam injection, previously used in chemistry studies,[11] to locally dope the plasma in a POS experiment with various species used to select emission lines for different spectroscopic observations. Unlike the previously reported method based on laser evaporation,[10] the present method allows for the injection of gaseous atoms for which the ionization times are sufficiently long to extend the measurement of line intensities and spectral profiles to a longer period during the switch operation. Evidently, line emission from ions produced due to the ionization of the injected atoms can also be used for various measurements, as is shown below.

In doping plasmas, it is required that the density of the doped atoms and their ions be sufficiently low so as to cause no significant modification of the properties of the plasma studied. In addition, in order to obtain local measurements, the doped region has to be significantly smaller than the plasma. This necessitates an injected gas beam that is satisfactorily collimated and that has a sufficiently sharp front, which allows for a short time delay between the gas injection and the application of the pulse that powers the device, thus minimizing gas scattering from electrode surfaces.

Our doping arrangement is presently used to investigate a planar-geometry POS with a peak current of 180 kA conducted by the plasma for 400–600 ns before the switch opens into an inductive load within less than 100 ns. The plasma used to prefill the POS region is generated by a flashboard plasma source[12] and is injected through slots in the anode into the 2.6-cm-wide interelectrode gap (along the $x$ direction, see Fig. 1), filling a region 14 cm wide (along the $y$ direction) and 8–10 cm long (along the $z$ direction). The flashboard consists of sixteen parallel chains, each with 12 flashover gaps, and is driven by a 2.8 $\mu$F capacitor charged to 35 kV.

The present POS experiment is designed for studying the interaction between the magnetic field due to the applied current and the plasma, with the aim of determining the magnetic field distribution with satisfactory spatial resolution using observations of line-emission Zeeman splitting.[13] For Zeeman splitting measurements, the Doppler broadening of the spectral lines should be minimized, which can be obtained by using emission from atoms (that are not accelerated by the magnetic fields). The plasma doping method was also utilized here to measure the velocities of various ions (HeII, NeII, ArIII, and XeII) by observing the line-emission Doppler shifts. Measurements for ions of different charges and masses can be used to study the details of the ion acceleration under the magnetic-field gradients and the ion collision-

[a)]Electronic mail: fnarad@plasma-gate.weizmann.ac.il
[b)]Present address: Queen's University, Belfast, Northern Ireland, UK.

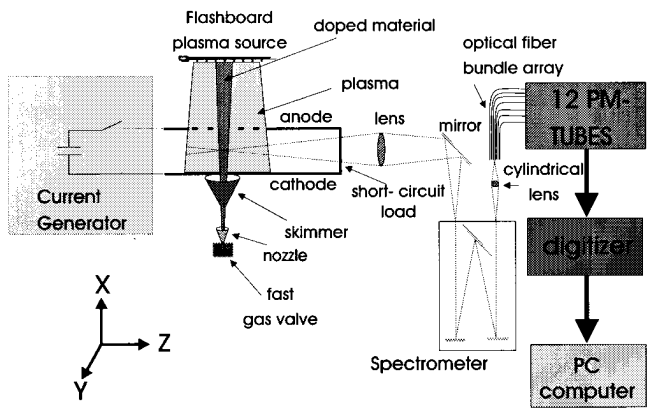

FIG. 1. The experimental system used to study the planar geometry POS. The interelectrode region is prefilled with plasma generated using a flashboard source and injected through slots in the anode. A fast gas valve, a nozzle, and a skimmer are used to inject a molecular beam for locally doping the plasma with various species. Light emitted from the doped column is collected by a lens parallel to the electrodes and focused onto a spectrometer, the output of which is focused by a cylindrical lens onto a rectangular fiber-bundle array which passes the light to 12 photomultiplier tubes.

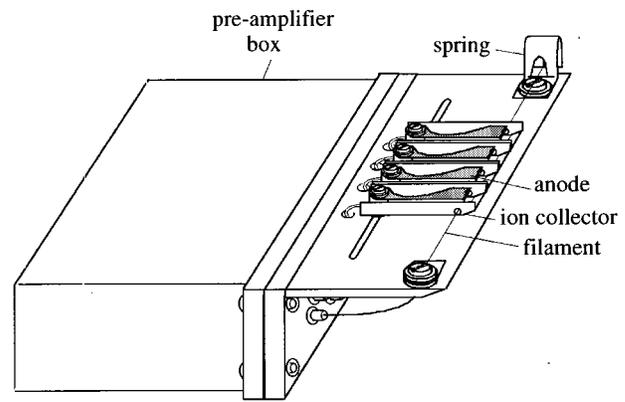

FIG. 2. The ionization-probe array. A single filament supplies thermal electrons to all four probes. Four anodes and ion collectors are used to collect the electrons and ions. A spring is used to maintain the tension in the filament during its heating. The preamplifiers are placed near the probes in order to obtain a satisfactory temporal response.

ality. Furthermore, the electron density and temperature can be studied from the temporal evolution of various emission line intensities.

An additional advantage in employing doping in spectroscopic measurements is that it allows for spectroscopic determination of the properties of plasmas from which no significant line emission is available, for example proton plasmas. In this case, line emission from the doped particles, excited by the plasma electrons, is used to obtain the plasma electron density and temperature.

## II. EXPERIMENTAL SETUP

The doping arrangement consists of a fast gas valve[14] and a 0.8-mm-diam nozzle. The valve is driven by a 2 $\mu$F, 6 kV capacitor discharged through a low inductance strip line. A skimmer is placed approximately 5 cm from the nozzle to further collimate the gas beam before it enters the POS region through a 1-cm-diam hole in the cathode. Skimmers with apertures in the range of 0.3–0.7 cm in diameter were used. The gas is supplied to the valve from an external reservoir at a pressure of 8 atm. The position of gas injection can be moved in two dimensions in order to dope the POS plasma in different regions in different experiments. In the present experiments, no differential pumping is required since we operate in single pulse mode. Figure 1 shows the experimental setup of the POS, the gas-doping arrangement, and the spectroscopic diagnostics. The gas-beam spatial profile and absolute density were diagnosed using both an array of ionization probes and spectroscopic methods. For the ionization probe measurements, we adopted a design similar to the one proposed by Schulz,[15] that allows for measurements in the range of $10^{-5}$–1 Torr. Figure 2 shows the scheme of our ionization probe array. A single 0.2-mm-diam tungsten wire was used as a filament to supply thermal electron emission to four ionizing cells. The dimensions of each cathode cell are $3\times0.7\times0.4$ cm and the anode plate is 0.5 cm wide and is placed 0.2 cm above the filament in order to maximize the electric field required for electron acceleration.

The structure of the ionization probes is so designed to maximize the free gas flow in order to minimize the perturbation to the original gas beam density. The ion collectors are bent at the front of the probe in order to shield the filament from direct impact by the gas beam that results in a drop in the electron emission during the measurement due to the cooling and the contamination of the hot filament. The side walls of the ion collectors separate the cells in order to minimize the ion current between the adjacent channels.

A voltage of 100 V relative to the grounded ion collectors is applied to the anodes and the filament is biased at 10 V. For some gases, with high ionization rates, the anodes were only charged to 60 V in order to obtain a larger linear detection range. The heating of the filament is controlled by a standard dc feed-back loop that stabilizes the total electron emission current at 200 $\mu$A. The ion currents from the four cathodes are amplified by preamplifiers placed near the probes. This allows the time response of the signals to be as low as 1 $\mu$s and the maximum cross talk between the different channels (that mainly stems from stray capacitance in the preamplifier circuit) to be about 5% at a frequency around 1 MHz.

The measurements with the ionization probe array were carried out with an experimental setup identical to that of the POS experiment except that the anode was removed to allow the probes to be inserted at different heights above the cathode.

The ionization probes were calibrated in dc mode for various operating gases using a known pressure. A 1 mTorr Mclaud gauge and a thermocouple were used as reference gauges and the pressure in the vacuum vessel was controlled by a needle valve during the calibration process. Figure 3 shows a calibration plot for He. The probes showed no saturation for pressures below 0.1 Torr (a gas density of $3.5\times10^{15}$ cm$^{-3}$) for all gases used.

The spatial density profile of the gas beam was also determined using line emission from the injected gas. To this end, the flashboard plasma was injected in order to allow for excitations of the doped atoms. The line emission intensity can be assumed to be proportional to the gas density, based on the uniformity of ±5% found for the flashboard plasma over a 1-cm-long region. Unlike the ionization probe measurement, in these measurements, the anode strips were not removed. This allows for the determination of the gas beam profile using the true experimental POS geometry, for which scattering from the anode strips is possible.

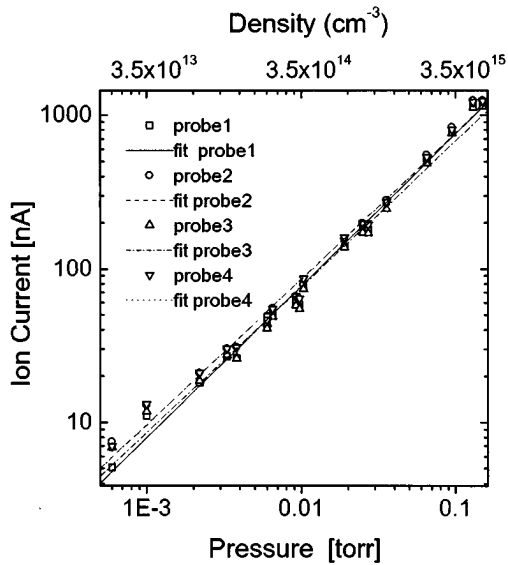

FIG. 3. Calibration curves of the four-channel ionization-probe array for helium. Both the horizontal and vertical errors are less than ±5% except at pressures below 0.01 Torr, where the horizontal error is ±0.5 mTorr because of the limitation of the Mclaud gauge.

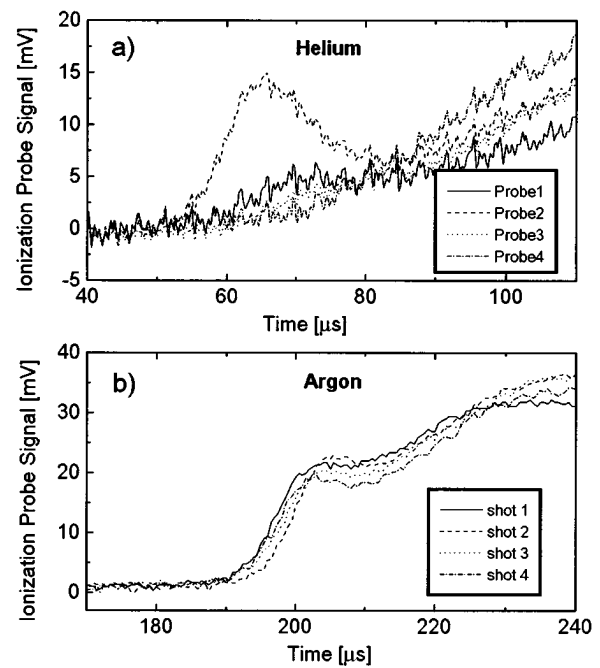

FIG. 4. (a) Traces of the four ionization-probe channels obtained using helium. The array was placed in the middle of the anode-cathode gap in front of the skimmer and a hole in the cathode. (b) Four traces of the central channel of the probe obtained in four successive injections of argon. Here, a 1 μs filter was used in order to smooth the traces.

The spectroscopic diagnostic system consists of a 1 m spectrometer for the range 2000–7500 Å, equipped with a 2400 grooves/mm grating. The light from the doped column is imaged onto the spectrometer using a lens and mirrors with a spatial resolution of 0.4 cm. A cylindrical lens images the light at the output of the spectrometer onto a rectangular fiber-bundle array, that transmits the light to a set of photomultiplier tubes. The output from the 12 photomultipliers is recorded by a multichannel digitizer, which gives the time dependent spectral line profile in a single discharge.

## III. MEASUREMENTS AND RESULTS

### A. Characterization of the doped-gas beam

In our measurements, the gas spatial density profile perpendicular to its propagation was determined for various gas species at various distances from the valve. Figure 4(a) shows traces of the four channels of the ionization probe array as a function of time for helium. The signals are obtained at a distance of 10 cm from the valve and are translated to density using the probe calibration. Probe No. 2 shows significantly larger signals than the other probes at $t = 55–80$ ns which indicates a narrow density profile at these times. Also, this figure shows the relatively fast rise of the gas density ($\approx 5$ μs), followed by a drop of the gas density within 15 μs. Figure 4(b) shows the signal of the central channel of the ionization probe array obtained in four different discharges using argon. Comparison to Fig. 4(a) shows the longer delay for the argon arrival and the longer pulse duration typical for heavier gases. For He and Ar, the gas velocity was found to be $(1.9\pm0.1)\times10^5$ and $(7\pm0.5)\times10^4$ cm/s, respectively. These velocities are used for selecting the time delay between the operation of the valve and the application of the current to the flashboard plasma source. The data in Fig. 4(b) also show that the reproducibility of the gas density is about ±8%.

An absolute density profile obtained using the ionization probe array with helium, is shown in Fig. 5(a) for three positions in the anode-cathode gap. The peak gas density varies from $0.8\times10^{14}$ cm$^{-3}$ near the anode to $1.7\times10^{14}$ cm$^{-3}$ near the cathode. In these experiments, a skimmer with an aperture of 0.3 cm was used, yielding column full width at half maximum (FWHM) between 0.7 and 0.8 cm. In part of these spectroscopic measurements, a skimmer with a larger aperture could be used to allow for higher light intensities at the expense of the spatial resolution. For example, Fig. 5(b) shows a profile of the HeI $3d(^1D)$ level population, obtained from the 6678 Å line intensity, using a skimmer with relatively large aperture ($\varnothing=0.7$ cm). The FWHM of the gas profiles were found to be 1.2–1.6 cm for this larger skimmer.

### B. Studies of the POS plasma

In this section we present examples of the use of the present doping arrangement in the investigation of the flashboard plasma and the POS (further results will be published elsewhere). The front of the plasma injected from the flashboard reaches the POS region about 600 ns after the operation of the flashboard and provides nearly no line emission, presumably since the front mainly consists of protons. The electron density of this plasma was determined from Stark broadening of doped-hydrogen lines and found to be ≈3 ×10$^{14}$ cm$^{-3}$. The electron temperature of this proton plasma was determined from the time behavior of line intensities of doped fast-ionizing species such as ArI, CI, and hydrogen. These line intensities are seen to drop as a result of ionization, and the ionization times thus obtained, together with the knowledge of the electron density and the use of collisional-radiative calculations,[16] allowed for the determination of the electron temperature, found to be 8–10 eV. This technique can also be used to study electron flow and proton plasmas between the POS and the load,[17,10] which is important for the understanding of the energy coupling to the load.

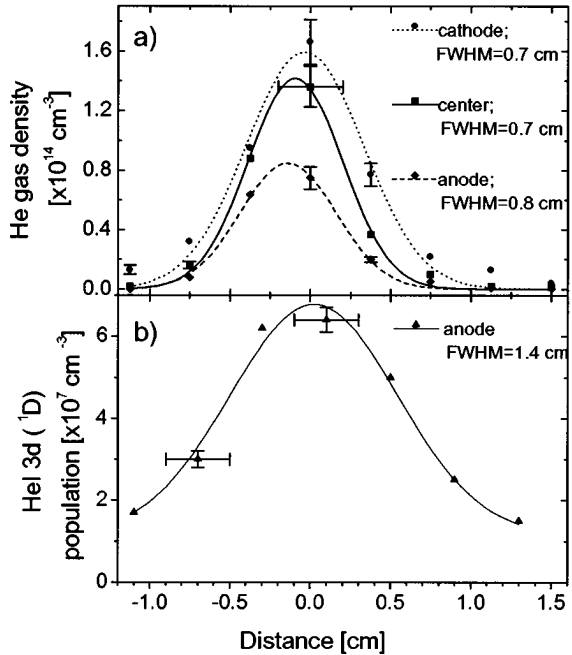

FIG. 5. (a) An absolute density profile for helium, measured at three positions in the anode-cathode gap, obtained using the ionization probe array. Eight data points are obtained by shifting the probes by 0.35 cm (half the distance between adjacent probes) in two consecutive measurements. The FWHM of the profiles are indicated. (b) An absolute density profile of the HeI $3d(^1D)$ level near the anode. Here, a skimmer with an aperture 0.7 cm in diameter was used. The uncertainties are indicated by the few error bars shown.

The electron density of the flashboard plasma, determined from Stark broadening of hydrogen lines, was seen to increase to $1 \times 10^{15}$ cm$^{-3}$ about 0.6 µs later. At this time, the fraction of the CIII and CIV ions in the plasma was found to be higher, i.e., particles heavier than protons reached the POS gap. Line emission of doped particles that ionize slowly such as HeI, ArII, and CII, and thus provide intense lines for a relatively long time, were used to determine the electron temperature at this time. The electron temperature at this time was found to be 4.5±0.5 eV, which is significantly lower than in the proton plasma.

The local axial ion velocities during the POS high-current pulse were determined from Doppler shifts of lines of various species injected into the plasma. Figure 6 shows the time dependent axial velocity of HeII, resulting from the ion-ization of doped helium, obtained from the 4686 Å line at 0.5 cm from the cathode at five axial positions. The traces are only shown for the periods in which the line intensity is high enough to yield a satisfactory accuracy. The velocities are seen to rise in time due to the acceleration under the gradient of the magnetic field. The start time of the He II acceleration at different $z$ locations allows the magnetic field average axial propagation velocity to be determined, found to be $(2.8 \pm 0.3) \times 10^7$ cm/s. The velocities at larger-$z$ locations rise to higher values since the velocity rise at these locations occurs at later times and the POS current rises in time. The relatively strong dependence of the ion velocities on the axial position demonstrates the need for local observations, made possible by the present doping technique.

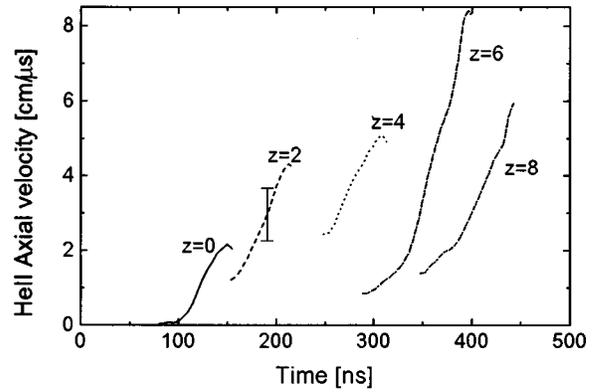

FIG. 6. Spatially resolved measurements of the HeII axial velocities in the POS near the cathode for five axial positions obtained from the Doppler shift of the $n = 4$ to $n = 3$ line. The times shown are relative to the beginning of the POS current pulse. The error bar shown indicates the typical uncertainty. The spatial resolution along the line of sight is ≈1.5 cm.

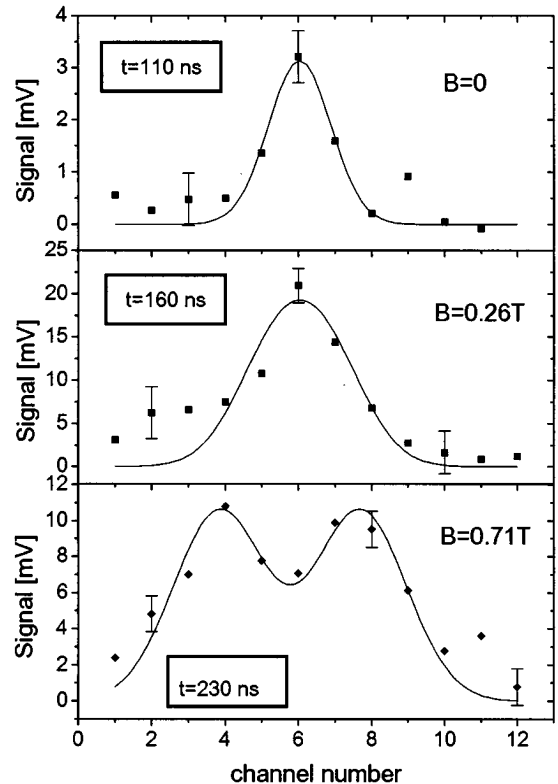

FIG. 7. Spectral profiles of the HeI 6678 Å line used to measure the magnetic field evolution during the POS operation near the cathode in the axial center of the POS, given for three times. Here, the spectral resolution is 0.07 Å/channel. The profile at $t = 110$ ns gives zero magnetic field, the one at 160 ns is broadened by the field, and the one at 230 ns shows splitting by the field. The profiles are fitted by two split Gaussians. The ion temperature is less than 1 eV, giving a profile that is dominated by the instrumental broadening. The spatial resolution along the line of sight is ≈1.5 cm, and the temporal resolution is ±20 ns at $t = 110$ and ±10 ns at later times.

Our doping technique also allowed for local determination of the magnetic field using line-emission Zeeman splitting. To this end, we observed the spectral profile of the helium atom 6678 Å line in the direction of the magnetic field (along the ''$Y$'' direction, see Fig. 1), thus only observing the $\sigma$ components. Figure 7 shows the line profile at $t = 110$, 160, and 230 ns in the middle of the $Y$ and $Z$ dimensions of the POS (see Fig. 1), and at 0.5 cm from the cathode. The line width of the He I line is dominated by the instrumental broadening since the Doppler broadening remains small throughout the pulse (the ion-atom collisionality is too low to cause a momentum transfer from the accelerated ions to the neutral particles). The line profile is fitted by two Gaussians (whose width is determined by the instrumental broadening) and the splitting between them yields the magnetic field. Zeeman splitting is clearly seen at $t = 230$ ns, giving $0.7 \pm 0.1$ T for the magnetic field amplitude. To within the uncertainties, the magnetic field propagation velocity obtained from the Zeeman splitting and the ion acceleration measurements are in agreement with each other. Further measurements using this method, including complete 3D mapping of the magnetic field, are currently in progress.

## ACKNOWLEDGMENTS

The authors would like to thank Y. Krasik, A. Weingarten, and K. Tsigutkin, for their useful suggestions and discussions, and P. Meiri for his technical assistance. Special thanks are due to W. R. Gentry from the University of Minnesota, Minneapolis, Minnesota for providing us with the gas valve and for his valuable advice. This work was supported by the Minerva foundation (Munich, Germany).